\documentclass[a4paper,12pt]{article}
\pdfoutput=1 % if your are submitting a pdflatex (i.e. if you have
\usepackage{jheppub} % for details on the use of the package, please
\usepackage[T1]{fontenc} % if needed

\usepackage{amsmath}
\usepackage{xcolor}

\def\be{\begin{equation}}
\def\ee{\end{equation}}
\def\n{\nonumber \\}

\def\bea{\begin{eqnarray}}
\def\eea{\end{eqnarray}}

\usepackage{amssymb}
\let\mathds\mathbb

\title{\boldmath Correlators for tensionless strings on ${\rm AdS}_3$ orbifolds}

\author[a]{Matthias R.\ Gaberdiel,}
\author[a,b]{and Bin Guo}

\affiliation[a]{Institut f\"ur Theoretische Physik, ETH Zurich, \\ CH-8093 Z\"urich, Switzerland}
\affiliation[b]{School of Physics, Central South University, \\ Changsha 410083, China}

\emailAdd{gaberdiel@itp.phys.ethz.ch}
\emailAdd{guobin@csu.edu.cn}

\abstract{The CFT dual of string theory on $({\rm AdS}_3 \times {\rm S}^3)/\mathbb{Z}_k\times \mathbb{T}^4$ is believed to be described by the subspace of the symmetric orbifold of $\mathbb{T}^4$ that comprises the low-lying excitations on top of a certain reference state. (This `non-perturbative' reference state lies in the twisted sector associated to the conjugacy class consisting of only $k$-cycles.) In a recent paper we confirmed this picture by analysing the worldsheet theory of the orbifold at the tensionless NS-NS point, and by showing that the perturbative worldsheet spectrum reproduces precisely the single particle excitations on top of this reference state. In this paper we explain that this identification also holds on the level of the correlators, at least to leading order in $1/N$.}

\begin{document} 
\maketitle
\flushbottom

\section{Introduction}

Recently, the AdS$_3$/CFT$_2$ duality for the case where the string background has one unit of NS-NS flux and the dual CFT is the symmetric orbifold of $\mathbb{T}^4$ was understood in detail \cite{Eberhardt:2018ouy,Eberhardt:2019ywk}. In particular, it was shown that the spectrum of the worldsheet theory (that can be described in the hybrid  description of \cite{Berkovits:1999im} in terms of a WZW model based on $\mathfrak{psu}(1,1|2)_1$) matches exactly that of the symmetric orbifold. Furthermore, the correlators of the symmetric orbifold could be reproduced from the string perspective by showing that the covering surface that appears as an auxiliary construction in the symmetric orbifold calculation \cite{Lunin:2000yv} can be identified with the worldsheet \cite{Eberhardt:2019ywk} --- thereby realising an old idea of \cite{Pakman:2009zz}.

In an independent development, an understanding has been developed  over the years for how symmetric orbifold results can be identified with supergravity calculations, see e.g.\ \cite{Lunin:2001jy,Giusto:2004id,Kanitscheider:2006zf,Giusto:2015dfa,Bena:2016ypk,Ganchev:2023sth}. Given that the background with one unit of NS-NS flux is far from the supergravity regime --- it describes the `tensionless' string for which the AdS space is of string scale --- it is interesting to explore to which extent this supergravity intuition is compatible with the explicit worldsheet results. A first step in this direction was recently undertaken in \cite{Gaberdiel:2023dxt} where the worldsheet theory for the case of an $\mathbb{Z}_k$ orbifold of ${\rm AdS}_3 \times {\rm S}^3$ (times $\mathbb{T}^4$) at the tensionless point was analysed. According to the expectations of \cite{Martinec:2001cf}, the dual CFT should be described by the subsector of the $N$'th symmetric orbifold of $\mathbb{T}^4$ (in the large $N$ limit) that consists of the low-lying excitations on top of a certain highly excited state, see eq.~(\ref{bg}) below, and the worldsheet analysis of \cite{Gaberdiel:2023dxt} confirmed exactly this prediction. 

This raises the question whether also the correlators of the two descriptions match, and this is what  we shall address in this paper. The situation is a little bit subtle since the  subspace of the symmetric orbifold that is dual to the $\mathbb{Z}_k$ worldsheet orbifold does not contain the actual symmetric orbifold vacuum, and there is therefore no canonical field-state correspondence that would allow us to identify the operators unambiguously from the spectrum. However, at least in the large $N$ limit and for the vertex operators associated to the untwisted sector states of the worldsheet orbifold, there is a natural guess for what the correct identification should be, and it takes the schematic form 
\be\label{aim0}
\langle\Psi|\prod_{i=1}^{n} V_i( x_{i})|\Psi\rangle_{\rm SymOrb}=\int_{\mathcal M} d\mu(z_i) \, \Big\langle \prod_{i=1}^{n} V_i(x_{i};z_i)\Big\rangle_{{\rm AdS}_3 \times {\rm S}^3/ \mathbb{Z}_k} \ ,
\ee
see also eq.~(\ref{aim}) below. The main result of the present paper is to confirm that this identification is indeed true. While the worldsheet calculation is relatively straightforward, the matching with the symmetric orbifold is somewhat non-trivial: in particular, it hinges on the fact that, to leading order in $\frac{1}{N}$,  the covering maps $\Gamma(z)$ that contribute to the symmetric orbifold calculation are of the form $\Gamma(z) = (\Gamma_0(z))^k$, where $\Gamma_0$ is another covering map. In fact, $\Gamma_0(z)$ turns out to be the covering map that controls the localisation of the worldsheet correlator.

Somewhat similar worldsheet calculations have been performed before in \cite{Bufalini:2022wyp,Bufalini:2022wzu} for the closely related JMaRT backgrounds \cite{Jejjala:2005yu} whose worldsheet description was developed in \cite{Martinec:2017ztd,Martinec:2018nco,Martinec:2019wzw}. However, these calculations were done at large level ${\sf k}$ (for which only unflowed representations of the worldsheet appear). This corresponds to the supergravity regime from the AdS perspective, but then the CFT dual is not the symmetric orbifold of $\mathbb{T}^4$, but rather the theory described in \cite{Eberhardt:2021vsx}, see also \cite{Eberhardt:2019qcl,Balthazar:2021xeh,Dei:2022pkr,Hikida:2023jyc,Yu:2024kxr,Yu:2025qnw}. On the other hand, here we concentrate on the tensionless case (at level ${\sf k}=1$), for which we have an exact duality to the (subsector of the) symmetric orbifold. 

\medskip

The paper is organised as follows. We review the salient features of the tensionless orbifold duality in Section~\ref{sec:review}. We also explain there some of the subtleties in relating the worldsheet operators to those of the dual CFT, see in particular Section~\ref{sec:vertex}, but argue that the natural identification at large $N$ is indeed eq.~(\ref{aim0}). We evaluate the relevant worldsheet expression in Section~\ref{sec:worldsheet}, and show that the right-hand-side of eq.~(\ref{aim0}) can be written as in eq.~(\ref{2.8}). In Section~\ref{sec correlator} we then reproduce this answer from the corresponding dual CFT calculation. The key step in that argument relies on a careful analysis of the covering maps that contribute at large $N$, see eq.~(\ref{key}) in Section~\ref{sec:3.1}. 
We can evaluate the corresponding conformal factor using a somewhat non-trivial identity of covering maps, see eq.~(\ref{i0}), and this reproduces precisely the worldsheet answer, see eq.~(\ref{bc}).  We close in Section~\ref{sec:conclusion} by mentioning interesting avenues for future research. There are three appendices in which we review (and derive) various properties of covering maps.

\section{Review of the duality for $({\rm AdS}_3\times {\rm S}^3)/ \mathbb{Z}_k \times \mathbb{T}^4$}\label{sec:review}
 
 Let us begin by reviewing tensionless string theory on $({\rm AdS}_3\times {\rm S}^3)/ \mathbb{Z}_k \times \mathbb{T}^4$ and its CFT dual as discussed in \cite{Martinec:2001cf,Gaberdiel:2023dxt}. The metric for ${\rm AdS}_3\times {\rm S}^3$ takes the form
\be\label{ads metric}
ds^2 = - (r^2 +1) dt^2 + \frac{dr^2}{r^2 +1} + r^2 dy^2 + d \theta^2 +\cos^2\theta d \psi^2 +\sin^2\theta d\phi^2 \ ,
\ee
where the coordinate identifications are
\be\label{ads p}
(y,\psi, \phi)\sim \left( y+2\pi, \psi, \phi  \right) \sim  \left( y, \psi +2\pi , \phi \right) \sim \left( y, \psi  , \phi +2\pi \right) \  .
\ee
Tensionless string theory on this background is described by a WZW model based on the superaffine algebra $\mathfrak{psu}(1,1|2)_1$, whose bosonic subalgebra $\mathfrak{sl}(2,\mathds{R})_1 \oplus \mathfrak{su}(2)_1$ corresponds to ${\rm AdS}_3$ and ${\rm S}^3$, respectively. We will denote the modes of $\mathfrak{sl}(2,\mathds{R})_1$ by $J^a_n$, while the generators of $ \mathfrak{su}(2)_1$ are $K^a_n$, where $a\in\{\pm,3\}$ and $n\in\mathbb{Z}$. 

We are interested in the $\mathbb{Z}_k$ orbifold of this background, where the orbifold acts on the spacetime coordinates as
\be\label{Zk iden}
(y,\psi, \phi)\sim \bigl( y+\tfrac{2\pi}{k}, \psi - \tfrac{2\pi(s+\bar s +1)}{k}, \phi + \tfrac{2\pi (s-\bar s)}{k} \bigr) \ ,
\ee
and $s$ and $\bar s$ are integers satisfying 
\be
\frac{s (s+1) - \bar{s} (\bar{s}+1)}{k} \in \mathbb{Z} \ .
\ee
The corresponding $\mathbb{Z}_k$ action on the worldsheet is then given by 
\be\label{worldsheet g Zk}
\mathbb{Z}_k: \quad g= e^{\frac{2\pi i}{k}\,  (J^3_0 -(2s+1) K^3_0)} \otimes e^{-\frac{2\pi i}{k} (\bar{J}^3_0 -(2\bar s+1) \bar{K}^3_0)}  \ .
\ee 
In the sector with spectral flow (or winding) $w$, the worldsheet excitations with excitation number $N^{\rm ws}$ (with respect to the torus modes) and $K^3_0$ charge $\delta_i$ have spacetime conformal dimension and $\mathfrak{su}(2)$ charge
\begin{align}
h =\, & J^3_0 =  \frac{N^{\rm ws}}{w} + \frac{w+2s+1}{2} + (2s+1)\sum_i \delta_i + w s(s+1) \ , \label{wss} 
\\
q=\, & K^3_0 =   \frac{w(2s+1)+1}{2}+\sum_i \delta_i  \ ,\label{wsj}
\end{align}
in the coordinates corresponding to (\ref{ads metric}), see \cite[eqs.~(4.14) and (4.15)]{Gaberdiel:2023dxt}. 

The formulae for the right-moving excitations are  mutatis mutandis the same. In the untwisted sector of the worldsheet orbifold $w\in\mathbb{N}$, and the twisted sectors correspond to the situation where $w$ is fractional in units of $\frac{1}{k}$. The four bosons are charge neutral ($\delta_i=0$), and only their negative modes act non-trivially. Of the fermions, two have $\delta_i=-\frac{1}{2}$ and two have $\delta_i=+\frac{1}{2}$. Only the zero modes of the fermions with $\delta_i=-\frac{1}{2}$, along with all the negative modes of all fermions, act non-trivially, see  \cite{Eberhardt:2018ouy,Gaberdiel:2023dxt} for details.

As usual, we shall work with Euclidean ${\rm AdS}_3 \times {\rm S}^3$, where the Euclidean time $t_{E}$ is related to the Lorentzian time by $t_{E}= it_{L}$; for ease of notation we will omit the subscript $E$ in the following. We furthermore write the ${\rm AdS}_3$ boundary coordinate as $x=e^{iy + t_L} = e^{iy - i t}$. For the following it will sometimes be convenient to use (rescaled) Ramond sector coordinates, which we will distinguish by the subscript $R$. They are related to the above coordinates via
\bea
&&t= \frac{t_R}{k} \ ,\quad y= \frac{y_R}{k} \ , \quad r=k r_R \ , \quad\theta=\theta_R \ ,\nonumber \\
&& \psi=\psi_R- \frac{s+\bar s+1}{k} y_R+\frac{s-\bar s}{k} t_R \ , \quad \phi=\phi_R+\frac{s-\bar s}{k}y_R-\frac{s+\bar s+1}{k} t_R  \ .
\label{Rcc1}
\eea
Here the shifts in $\psi$ and $\phi$ are generated by the operators $K^3_0+\bar K^3_0=i\partial_\phi$ and $K^3_0-\bar K^3_0=-i\partial_\psi$ through
\be\label{xRact}
e^{-i\left(\frac{s-\bar s}{k}y_R-\frac{s+\bar s+1}{k}t_R\right)(K^3_0+\bar K^3_0)+i\left(-\frac{s+\bar s+1}{k}y_R+\frac{s-\bar s}{k}t_R\right)(K^3_0-\bar K^3_0)}
 = x_R^{-\frac{2s+1}{k}K^3_0} \bar x_R^{-\frac{2\bar s+1}{k}\bar K^3_0} \ ,
\ee
where
\be\label{xR}
x_R = e^{i y_R - i t_R} \ , \qquad \bar{x}_R = e^{-i y_R - i t_R} \ .
\ee
Then the identification (\ref{Zk iden}) becomes the asymptotic ${\rm AdS}_3\times {\rm S}^3$ identification, analogous to that in (\ref{ads p})
\be
(y_R, \psi_R, \phi_R)\sim (y_R+2\pi, \psi_R, \phi_R)\sim (y_R, \psi_R+2\pi, \phi_R)\sim (y_R, \psi_R, \phi_R+2\pi) \ .
\ee
In this description the charges of a string excitation are given by
\begin{align}\label{dc}
h^R=J_0^{3\,R}&= \frac{1}{k} J_0^3- \frac{2 s+1}{k} K_0^3  = \frac{N^{\rm ws}}{wk}- \frac{ws(s+1)}{k}
\ , \nonumber \\
q^R=K_0^{3\,R} &= K_0^3= \frac{w(2s+1)+1}{2}+\sum_i\delta_i \ .
\end{align}
From the perspective of the dual CFT, the transition between these two coordinates is described by ${\cal N}=4$ spectral flow, see the discussion in \cite[Section~2.4]{Gaberdiel:2023dxt}.

\subsection{The boundary CFT dual}

The dual CFT description was already proposed some time ago in \cite{Martinec:2001cf}: the physical states of the worldsheet theory correspond to single particle excitations in the symmetric orbifold, but now not relative to the orbifold ground state, but rather with respect to a certain reference state, see eq.~(\ref{bg}) below. This fits with the general expectations from \cite{Eberhardt:2020bgq,Eberhardt:2021jvj}, and could be quantitatively confirmed in \cite{Gaberdiel:2023dxt}. 

More specifically,  the relevant reference state (in the Ramond sector) is \cite[eq.~(4.18)]{Gaberdiel:2023dxt}
\be\label{bg}
|\Psi\rangle=\big(|\Psi_k\rangle_{s,\bar s}\big)^{N/k} \ ,
\ee
where
\begin{align}\label{bs a}
|\Psi_k\rangle_{s,\bar s} = \ & d^{-+}_{-\frac{s}{k}}d^{--}_{-\frac{s}{k}}\dots 
d^{-+}_{-\frac{2}{k}}d^{--}_{-\frac{2}{k}}d^{-+}_{-\frac{1}{k}}d^{--}_{-\frac{1}{k}} \n
&\ \times \bar d^{-+}_{-\frac{\bar s}{k}}\bar d^{--}_{-\frac{\bar s}{k}}\dots 
\bar d^{-+}_{-\frac{2}{k}}\bar d^{--}_{-\frac{2}{k}}\bar d^{-+}_{-\frac{1}{k}}\bar d^{--}_{-\frac{1}{k}}
\ |0^{--}_k\rangle_{\rm R} \ ,
\end{align}
and we are using the same conventions as in  \cite{Gaberdiel:2023dxt}, see in particular \cite[Section~2.3]{Gaberdiel:2023dxt}. Here, $|0^{--}_k\rangle_{\rm R}$ denotes the Ramond ground state in the $k$-cycle twisted sector, carrying charges $h^R=\bar h^R=\frac{k}{4},\ q^R=\bar q^R=-\frac{1}{2}$. 
The left-moving fermionic modes $d^{\alpha A}_{-\frac{n}{k}}$ have charges $h^R=\frac{n}{k}$ and $ q^R=\frac{\alpha}{2}$, where $\alpha=\pm$. 
Similarly, the right-moving fermionic modes $\bar d^{\bar \alpha A}_{-\frac{n}{k}}$  carry charges $\bar h^R=\frac{n}{k}$ and $\bar q^R=\frac{\bar \alpha}{2}$ again with $\bar \alpha=\pm$. The label $A=\pm$ distinguishes between the two species of fermions. 

It was shown in \cite{Gaberdiel:2023dxt} that the states in the $w$ spectrally flowed sector on the worldsheet (with $w\in \frac{1}{k} \mathbb{N}$) correspond to the excitations where a cycle of length $wk$ is created --- the remaining cycles all remain $k$ cycles and are described by the state in (\ref{bs a}). Indeed, it was shown there that the worldsheet excitation spectrum (\ref{dc}) matches exactly that in the dual CFT.

\subsection{The string vertex operator}\label{sec:vertex}

In this paper we want to relate the correlation functions of the two descriptions to one another, and for that we need to identify the dual CFT operators. This is now a bit non-trivial since the dual CFT is a subsector of the symmetric orbifold, namely the single-cycle excitations on the top of the non-trivial `vacuum' (\ref{bg}), and this sector does not contain the original symmetric orbifold vacuum. As a consequence, the usual field-state correspondence is not directly applicable. In addition, the definition of correlation functions for twisted sector states is delicate since the orbifold action (\ref{worldsheet g Zk}) involves $J^3_0$, and this is to be identified with the $L^{\rm CFT}_0$ operator of the dual CFT. As a consequence, we can only insert twisted sector operators at the fixed points of $L^{\rm CFT}_0$, i.e.\ at the north and south pole of the sphere at infinity  --- in terms of the variables of eq.~(\ref{ads metric}) this corresponds to $t=\pm \infty$, while in the usual CFT variables this is described by $x=0$ and $x=\infty$. Thus the most general correlation function we can consider consists of at most two twisted sector vertex operator inserted at $t=\pm\infty$, describing a twisted  `in'- and `out'-state, with all the other vertex operators that are inserted at $x\neq 0,\infty$ coming from the untwisted sector. 

In the following we shall concentrate on the simplest case where there are no twisted sector states at $x=0$ and $x=\infty$. Then all vertex operators are from the untwisted sector, i.e.\ are associated with 
spectral flow $w\in\mathbb{N}$. In order to identify the dictionary between the wordsheet and dual CFT operators we note that, when applied to the non-trivial vacuum (\ref{bg}), the twist operator $\sigma_w$ of order $w$ will generate (among others) a state that contains one $wk$ cycle, with all other cycles being $k$-cycles. In fact, as we shall argue, this is the dominant contribution at large $N$, and we therefore propose that, at least to leading order in $\frac{1}{N}$,\footnote{This dictionary may have to be corrected at subleading orders in $\frac{1}{N}$.}  the vertex operator associated to a state in the $w$-spectrally flowed sector on the worldsheet is a CFT operator from the $w$-cycle twisted sector. Obviously, this is also natural from the perspective of the original duality of \cite{Gaberdiel:2018rqv,Eberhardt:2018ouy}, where this was the correct prescription.\footnote{In that case, this identification was unambiguous using the usual state-field correspondence where one identifies a state with the operator that creates the state in question from the vacuum.}

It is the  aim of this paper is to show that this is indeed the correct identification, i.e.\ that the correlators calculated in the symmetric orbifold are correctly reproduced from the worldsheet in the large $N$ limit. We should stress that this does not follow automatically from the original analysis in \cite{Eberhardt:2019ywk} since the symmetric orbifold calculation differs from what was considered there: in particular, the symmetric orbifold correlators must now be calculated with respect to the reference state $|\Psi\rangle$ of eq.~(\ref{bg}), and thus the expected correspondence is schematically, see eq.~(\ref{aim0}) in the introduction
\be\label{aim}
\langle\Psi|\prod_{i=1}^{n} V_i( x_{i})|\Psi\rangle_{\rm SymOrb}=\int_{\mathcal M} d\mu(z_i) \, \Big\langle \prod_{i=1}^{n} V_i(x_{i};z_i)\Big\rangle_{{\rm AdS}_3 \times {\rm S}^3/ \mathbb{Z}_k} \ ,
\ee
where the $z_i$ are the coordinates on the worldsheet, and the worldsheet integral is done over the relevant moduli space ${\cal M}$.

In the following we shall first explain how to calculate the right-hand-side of (\ref{aim}), see Section~\ref{sec:worldsheet}. In Section~\ref{sec correlator} we shall then discuss the left-hand-side, and explain how to evaluate it in the large $N$ limit. As we shall show, see eq.~(\ref{bc}), this then reproduces exactly the worldsheet answer.

\section{Worldsheet correlators}\label{sec:worldsheet}

In this section, we study the string tree-level worldsheet correlators of the orbifold theory. For the case without any orbifold, the world-sheet computation reduces to 
\be
\int_{\mathcal M}d\mu(z_i) \, \Big\langle \prod_{i=1}^{n} V_i(x_i;z_i)\Big\rangle_{{\rm AdS}_3 \times {\rm S}^3} \ ,
\ee
where $z_i$ are the worldsheet coordinates, and the integration is performed over the moduli space of the $n$-point punctured sphere, i.e.\ it amounts to doing $n-4$ cross-ratio integrals. 

For the case of the orbifold $({\rm AdS}_3\times {\rm S}^3)/ \mathbb{Z}_k$ we will only consider untwisted sector operators, and we need to make them orbifold invariant. This is to say, the worldsheet correlators of the orbifold invariant vertex operators are of the form 
\be\label{m}
\Big\langle \prod_{i=1}^{n} V_i(x_i;z_i)\Big\rangle_{{\rm AdS}_3 \times {\rm S}^3/ \mathbb{Z}_k}=\Bigg\langle \prod_{i=1}^{n} \, \, \Bigl[ \sum_{\ell=0}^{k-1} \frac{1}{k}g^\ell \, V_i(x_i;z_{i})g^{-\ell} \Bigr]\Bigg\rangle_{{\rm AdS}_3 \times {\rm S}^3} \ ,
\ee
where the orbifold action $g$ is given by (\ref{worldsheet g Zk}).
Since the orbifold action involves $J^3_0$, which acts on the spacetime $x_i$ coordinate, the orbifold invariance condition implies that we need to sum over (in general) $k$ images of $x_i$. If $V_{i}$ carries the charges $q_i$ and $\bar q_i$ under $K^3_0$ and $\bar K^3_0$, respectively, we pick up phase factors from the $K^3_0$ and $\bar{K}^3_0$ exponential in (\ref{worldsheet g Zk}), and the right-hand-side of (\ref{m}) becomes 
\begin{align}\label{image0}
& \Big\langle \prod_{i=1}^{n} V_i(x_i;z_i)\Big\rangle_{{\rm AdS}_3 \times {\rm S}^3/ \mathbb{Z}_k} \n
& \qquad = \frac{1}{k^n}\Bigg\langle \prod_{i=1}^{n} \sum_{\ell=0}^{k-1} e^{\frac{2\pi i}{k}\ell\left(h_i-\bar h_i -(2s+1)q_i + (2\bar s+1)\bar q_i \right)}V_i(e^{\frac{2\pi i }{k}\ell}x_i;z_{i})\Bigg\rangle_{{\rm AdS}_3 \times {\rm S}^3} \ .
\end{align}
For the comparison to the dual CFT it is convenient to work in the (rescaled) Ramond sector coordinates, which are related to the coordinates above by the coordinate transformation (\ref{Rcc1}). Note that $K^3_0$ and $\bar{K}^3_0$ also act non-trivially on these $x_R$ coordinates, see eq.~(\ref{xRact}), and we therefore get additional factors of 
\be
x_{i,R}^{-(2s+1)q_i \frac{1}{k}} \, {\bar x}_{i,R}^{- (2\bar s+1)\bar q_i \frac{1}{k}} \ . 
\ee
In addition, the rescaling $t= \frac{t_R}{k}$ and $y= \frac{y_R}{k}$ of eq.~(\ref{Rcc1}) corresponds to the transformation
\be
x = x_R^{\frac{1}{k}} \ ,  \qquad \hbox{with} \qquad 
 \frac{\partial x}{\partial x_R} = \frac{1}{k} x_R^{\frac{1}{k} -1} \ , 
\ee
and if $V_i$ is primary of conformal dimension $h$ (as we shall assume), this leads to the conformal factor 
\be\label{ct}
\Big(\frac{\partial x}{\partial x_R}\Big)^h=k^{-h}x_R^{(\frac{1}{k}-1)h} \ , 
\ee
and similarly for the right-movers. Altogether we therefore find, see also \cite{Bufalini:2022wyp,Bufalini:2022wzu}
\begin{align}\label{wc}
&\Big\langle \prod_{i=1}^{n} V_i(x_{i,R};z_i)\Big\rangle_{{\rm AdS}_3 \times {\rm S}^3/ \mathbb{Z}_k}\n
& = \frac{1}{k^n}\Big\langle \prod_{i=1}^{n} \sum_{\ell=0}^{k-1} k^{-h_i-\bar h_i} 
e^{\frac{2\pi i}{k}\ell \left(h_i-\bar h_i-(2s+1)q_i + (2\bar s+1)\bar q_i \right)} \n
&\hspace{2cm}x_{i,R}^{-(2s+1)q_i \frac{1}{k}}
{\bar x}_{i,R}^{- (2\bar s+1)\bar q_i \frac{1}{k}}
{x_{i,R}}^{(\frac{1}{k}-1)h_i} {\bar x_{i,R}}^{(\frac{1}{k}-1)\bar h_i}
V_i(e^{\frac{2\pi i }{k}\ell }x_{i,R}^{\frac{1}{k}};z_{i})\Big\rangle_{{\rm AdS}_3 \times {\rm S}^3} \ . 
\end{align}
Since this is now a standard ${\rm AdS}_3 \times {\rm S}^3$ correlator, the old localisation argument of \cite{Eberhardt:2019ywk} resp.\ \cite{Dei:2020zui,Eberhardt:2025sbi} applies, and it follows that after integration over ${\cal M}$ we obtain  
\begin{align}
& \int_{\cal M} d\mu(z_i) \, \Big\langle \prod_{i=1}^{n} V_i(x_i;z_i)\Big\rangle_{{\rm AdS}_3 \times {\rm S}^3/ \mathbb{Z}_k} \n
   & =     \frac{1}{k^n} \sum_{\ell_1=0}^{k-1} \cdots \sum_{\ell_n=0}^{k-1} \Biggl[ \prod_{i=1}^{n} 
   (k x_{i,R})^{-h_i}
   (k \bar x_{i,R})^{-\bar h_i}
(e^{\frac{2\pi i}{k}\ell_i}x_{i,R}^{\frac{1}{k}})^{ h_i-(2s+1)q_i} 
(e^{-\frac{2\pi i}{k}\ell_i}{\bar x}_{i,R}^{\frac{1}{k}})^{ \bar h_i-(2\bar s+1)\bar q_i} 
\Biggr]
\n
&\hspace{1.5cm}  \times
\big\langle\Omega\big|
 V_1(e^{\frac{2\pi i }{k}\ell_1}x_{1,R}^{\frac{1}{k}}) \cdots V_n(e^{\frac{2\pi i }{k}\ell_n}x_{n,R}^{\frac{1}{k}} )\big|\Omega\big\rangle_{\rm SymOrb}  \ . \label{2.8}
\end{align}
This is the result we will compare with the dual CFT computation.

\section{The dual CFT correlators}\label{sec correlator}

In this section we will show that the dual CFT correlator evaluated with respect to the background state $|\Psi\rangle$, see eq.~(\ref{bg}), coincides with the right-hand-side of eq.~(\ref{2.8}). Our calculation proceeds in three  steps.

\subsection{The relevant covering maps}\label{sec:3.1}

The first key observation is that, in the large $N$ limit --- this corresponds to the sphere diagram from the worldsheet prespective --- the leading contribution on the dual CFT side arises from spherical covering maps that can be written as 
\be\label{key}
x_R= \Gamma(z) = \Gamma_0^k(z) \ , 
\ee
where $\Gamma_0$ is also a spherical covering map. Diagramatically this simply means that all intermediate states in the symmetric orbifold correlator come from sectors where all cycle lengths are integer multiples of $k$. 

The argument to see this is actually relatively straightforward. As we explain in Appendix~\ref{sec sc} we can write any spherical covering map as 
\be\label{map0}
x=\Gamma(z)=\hat{b} \prod_{j=1}^{m}(z-r_j)^{N_j}  \ .
\ee
For the case at hand, all cycles at $x=0$ and $x=\infty$ are $k$-cycles, and thus all $N_j$ must be either $N_j=k$ or $N_j=-k$. But then it is immediate from (\ref{map0}) that we can write 
\be
\Gamma(z) = \bigl( \Gamma_0(z) \bigr)^k \ , \qquad \hbox{with} \qquad 
\Gamma_0(z)=b\prod_{j=1}^{m}(z-r_j)^{S_j} \ , \qquad S_j \in \{ \pm 1\} \ ,
\ee
where $b^k = \hat{b}$. 

While the above argument is conclusive, it is also instructive to understand the result using explicit permutation group considerations; this will be described in Appendix~\ref{app:B}.

\subsection{Determining the correlator from the covering map}\label{sec:4.1}

The second step consists of using the explicit form of the above covering map to calculate the dual CFT correlator. We begin by recalling that, as explained in \cite{Lunin:2000yv,Dei:2019iym}, see also Appendix~\ref{app c} for some slight generalisation, the correlator is simply given by, see eq.~(\ref{vca}) in Appendix~\ref{app c}
\be\label{vc}
\big\langle  \prod_{i=1}^{n} V_i( x_{i,R}) \big\rangle = C_{\Gamma}\, \big\langle  \prod_{i=1}^{n} \tilde V_i( z_{i})  \big\rangle \ ,
\ee
where the operators with a tilde describe the corresponding operators on the covering surface, and the constant $C_\Gamma$ equals, see eq.~(\ref{fr0a})
\begin{align}\label{fr0}
C_{\Gamma}
= |\tilde N|^{\frac{1}{2}(\tilde N+1)}|b|^{\frac{1}{2}(1-\frac{1}{\tilde N})}\Big(\prod_{j=1}^m |N_j|^{-\frac{1}{2}(N_j+1)} |c_j|^{-\frac{1}{2}(1-\frac{1}{N_j})}\Big)
\Big(\prod_{i=1}^n w_i^{-\frac{1}{2}(w_i+1)} |a_i|^{-\frac{1}{2}(1-\frac{1}{w_i})}\Big) \ .
\end{align}
Here the different parameters depend only on the covering map and are explained in Appendix~\ref{sec sc}. The main step is therefore to determine the prefactor $C_\Gamma$. 

We begin by noting that for the covering map $x_R=\Gamma_0^k(z)$, the local behavior similar to eqs.~(\ref{lb1}),  (\ref{lb2}), and (\ref{xj0}) becomes
\begin{align}
x_R &\approx c^k_j(z-r_j)^{kN_j} \ , \n
x_R &\approx b^k z^{k\tilde N} \ ,\n
x_R &\approx (x_i+ a_i(z-z_i)^{w_i})^k 
\approx  x^k_i+ kx^{k-1}_ia_i(z-z_i)^{w_i} \ ,
\end{align}
where $a_i$, $b$, and $c$ correspond to the corresponding quantities for the covering map $\Gamma_0(z)$. Thus the correlator $C_{\Gamma_0^k}$ for the covering map $x_R=\Gamma_0^k(z)$ can be obtained by making the following substitutions in the expression for $C_{\Gamma}$ as given in (\ref{fr0})
\be
N_j\to kN_j \ , \ \tilde N\to k\tilde N \ ,  \  c_j\to c_j^k \ ,\  b\to b^k \ ,\  a_i \to kx_i^{k-1}a_i \ .
\ee
This leads to the contribution
\begin{align}\label{fr01}
C_{\Gamma_0^k}
=\ &|k\tilde{N}|^{\frac{1}{2}(k\tilde{N}+1)}|b|^{\frac{1}{2}(k-\frac{1}{N})}\Big(\prod_{j=1}^m |kN_j|^{-\frac{1}{2}(kN_j+1)} |c_j|^{-\frac{1}{2}(k-\frac{1}{N_j})}\Big) \n
&\qquad\qquad \times\Big(\prod_{i=1}^n w_i^{-\frac{1}{2}(w_i+1)} |kx_i^{k-1}a_i|^{-\frac{1}{2}(1-\frac{1}{w_i})}\Big) \ .
\end{align}
Using identity (\ref{i0}) from Appendix~\ref{sec sc}, this can then be simplified to 
\begin{align}\label{Ck}
C_{\Gamma_0^k}
=\Big(\prod_{i=1}^n \big|k x^{k-1}_i\big|^{-\frac{1}{2}w_i+\frac{1}{2w_i}} \Big) \, C_{\Gamma_0} 
= \Big(\prod_{i=1}^n \Big|\frac{\partial  x_{i,R}}{\partial x_i}\Big|^{-h_i-\bar h_i} \Big)\, C_{\Gamma_0} \ ,
\end{align}
where $C_{\Gamma_0}$ is given by (\ref{fr0}). In the second equality, we have used the fact that the dimension of the twist operator of order $w_i$ is given by $h_i=\bar h_i = \frac{w_i}{4}-\frac{1}{4w_i}$. %This result holds for any $N_i$, not just $N_i=\pm 1$. 
Note that if we consider a descendant of the twist sector ground state, the argument works similarly, and $(h_i,\bar{h}_i)$ continue to be the conformal dimensions of $V_i$.

Finally, we note that, given the location $x_{i,R}$ of the operator $V_{i}(x_{i,R})$, there will be $k$ preimages $x^{(j)}_{i}$ that are mapped to $x_{i,R}$ as $x_{i,R}=(x^{(j)}_{i})^k$.  To express $C_{\Gamma_0^k}$ in terms of $C_{\Gamma_0}$ via (\ref{Ck}), we need to sum over these different preimages. Since each such operator picks up the conformal factor as shown in eq.~(\ref{Ck}), this therefore leads to 
\begin{align}\label{V cover}
V_i(x_{i,R}) &\to \frac{1}{k}\sum_{\ell_i=0}^{k-1}
\Big(\frac{\partial x^{(\ell_i)}_i}{\partial x_{i,R}}\Big)^{h_i} 
\Big(\frac{\partial \bar x^{(\ell_i)}_i}{\partial \bar x_{i,R}}\Big)^{\bar h_i} 
V_i(e^{\frac{2\pi i }{k}\ell_i}x_{i,R}^{\frac{1}{k}}) \n
&= \, \frac{1}{k}\sum_{\ell_i=0}^{k-1} k^{-h_i-\bar h_i}e^{\frac{2\pi i }{k}\ell_i(h_i-\bar h_i)}x_{i,R}^{(\frac{1}{k}-1)h_i}\bar x_{i,R}^{(\frac{1}{k}-1)\bar h_i}V_i(e^{\frac{2\pi i }{k}\ell_i}x_{i,R}^{\frac{1}{k}}) \ ,
\end{align}
where the prefactor $1/k$ ensures the proper normalisation of the operator, as obtained by considering the two point function. 

This accounts for all the vertex operators, but we also need to understand what happens to the 
background state $|\Psi\rangle$ of eq.~(\ref{bg}). We claim that it becomes 
\begin{align}\label{1s}
|\Psi_k\rangle_{s,\bar s} \to \  & d^{-+}_{-s}d^{--}_{-s}
\dots 
d^{-+}_{-2}d^{--}_{-2}d^{-+}_{-1}d^{--}_{-1} \n
&\ \times \bar d^{-+}_{-\bar s}\bar d^{--}_{-\bar s}\dots 
\bar d^{-+}_{-2}\bar d^{--}_{-2}\bar d^{-+}_{-1}\bar d^{--}_{-1}
\ |0^{--}_1\rangle_{\rm R} \ ,
\end{align}
since under the lift to the covering surface each $k$-cycle is replaced by a trivial cycle. Here, $|0^{--}_1\rangle_{\rm R}$ is the Ramond ground state of a single $\mathbb{T}^4$, for which the conformal dimensions equal $h^R=\bar h^R=\frac{1}{4}$, and we consider the state with $q^R=\bar q^R=-\frac{1}{2}$. The fact that this is the correct lift of $|\Psi\rangle$  was shown in \cite{Lunin:2001pw}. 

The third and final step then consists of applying an additional ${\cal N}=4$ spectral flow with parameters $\alpha'=2s+1$ and $\bar\alpha'=2\bar s+1$ to the whole correlator. Under this transformation a vertex operator $V_i$ of charge $(q,\bar{q})$ transforms as 
\be\label{st}
V_i(x) \to x^{-\alpha q}\, \bar{x}^{-\bar{\alpha} \bar{q}} \, V_i(x) \ ,
\ee
and hence each term on the right-hand side of (\ref{V cover}) becomes
\begin{align}\label{sf1}
V_i(e^{\frac{2\pi i }{k}\ell_i}x_{i,R}^{\frac{1}{k}}) \to 
(e^{\frac{2\pi i }{k}\ell_i}x_{i,R}^{\frac{1}{k}})^{-(2s+1)q_i}
(e^{-\frac{2\pi i }{k}\ell_i}\bar x_{i,R}^{\frac{1}{k}})^{-(2\bar s+1)\bar q_i} 
V_i(e^{\frac{2\pi i }{k}\ell_i}x_{i,R}^{\frac{1}{k}}) \ .
\end{align}
On the other hand, this ${\cal N}=4$ spectral flow maps the state (\ref{1s}) to the NS sector vacuum, $
|0_1\rangle_{\rm NS}$. Collecting all the different pieces together, the dual CFT correlator thus becomes 
\begin{align}\label{bc}
&\langle\Psi|\prod_{i=1}^{n} V_i( x_{i,R})|\Psi\rangle_{\rm SymOrb}\n 
& =     \frac{1}{k^n} \sum_{\ell_1=0}^{k-1} \cdots \sum_{\ell_n=0}^{k-1} \Biggl[ \prod_{i=1}^{n} 
   (k x_{i,R})^{-h_i}
   (k \bar x_{i,R})^{-\bar h_i}
(e^{\frac{2\pi i}{k}\ell_i}x_{i,R}^{\frac{1}{k}})^{ h_i-(2s+1)q_i} 
(e^{-\frac{2\pi i}{k}\ell_i}{\bar x}_{i,R}^{\frac{1}{k}})^{ \bar h_i-(2\bar s+1)\bar q_i} 
\Biggr]
\n
&\hspace{1.5cm}  \times
\big\langle\Omega\big|
 V_1(e^{\frac{2\pi i }{k}\ell_1}x_{1,R}^{\frac{1}{k}}) \cdots V_n(e^{\frac{2\pi i }{k}\ell_n}x_{n,R}^{\frac{1}{k}} )\big|\Omega\big\rangle_{\rm SymOrb}  \ ,
\end{align}
which therefore reproduces exactly eq.~(\ref{2.8}). 

\subsection{Deforming away from the orbifold point}

In the previous section we have shown that the boundary CFT correlator can be written, in the large $N$ limit, as a sum over images, see eq.~(\ref{bc}). A similar structure also appeared in the analysis of \cite{Bufalini:2022wyp,Bufalini:2022wzu} in which they considered the NS-NS background with larger flux ${\sf k}>1$. That background is believed to be dual to the more complicated (deformed) symmetric orbifold of \cite{Eberhardt:2021vsx} that involves an ${\cal N}=4$ Liouville factor in addition to the $\mathbb{T}^4$, see also \cite{Eberhardt:2019qcl,Balthazar:2021xeh,Yu:2024kxr,Yu:2025qnw}.\footnote{Note that one cannot change the value of ${\sf k}$ by an exactly marginal deformation of the symmetric orbifold. These backgrounds are therefore `far away' in moduli space from the tensionless case of ${\sf k}=1$.} Furthermore, the correlators at ${\sf k}>1$ are not believed to be localising as in \cite{Eberhardt:2025sbi}, see e.g.\ \cite{Dei:2021xgh,Dei:2021yom}, and thus the dual CFT cannot literally be a symmetric orbifold --- as is indeed the case for the proposal of \cite{Eberhardt:2021vsx}.

While it is therefore difficult to repeat the above analysis for the case with ${\sf k}>1$, we can generalise our analysis to the theories that are related to the tensionless orbifold by an exactly marginal operator from the $2$-cycle twisted sector,
\be\label{D}
\lambda \int d^2 x \, D(x,\bar x) \ ,
\ee
where $D$ has conformal dimensions $(h,\bar h)=(1,1)$ and is charge-neutral. (From the AdS perspective, these operators switch on R-R flux, as was recently confirmed by a direct perturbative analysis in \cite{Gaberdiel:2023lco,Gaberdiel:2025smz}.) Since the deformation is charge-neutral and marginal, it is invariant under the transformations (\ref{V cover}) and (\ref{sf1}). Accordingly,   at $m$'th order in perturbation theory, (\ref{bc}) generalises to
\begin{align}\label{bc1}
&\langle\Psi|\prod_{i=1}^{n} V_i( x_{i,R})\Big(\lambda \int d^2 x_R \, D(x_R,\bar x_R)\Big)^m |\Psi\rangle_{\rm SymOrb}\n 
 & =     \frac{1}{k^n} \sum_{\ell_1=0}^{k-1} \cdots \sum_{\ell_n=0}^{k-1} \Biggl[ \prod_{i=1}^{n} 
   (k x_{i,R})^{-h_i}
   (k \bar x_{i,R})^{-\bar h_i}
(e^{\frac{2\pi i}{k}\ell_i}x_{i,R}^{\frac{1}{k}})^{ h_i-(2s+1)q_i} 
(e^{-\frac{2\pi i}{k}\ell_i}{\bar x}_{i,R}^{\frac{1}{k}})^{ \bar h_i-(2\bar s+1)\bar q_i} 
\Biggr]
\n
& \quad \ \times
\big\langle\Omega\big|
 V_1(e^{\frac{2\pi i }{k}\ell_1}x_{1,R}^{\frac{1}{k}}) \cdots V_n(e^{\frac{2\pi i }{k}\ell_n}x_{n,R}^{\frac{1}{k}} )\Big(\lambda \int d^2 x_R \, D(x_R,\bar x_R)\Big)^m\big|\Omega\big\rangle_{\rm SymOrb} \ ,
\end{align}
in the large $N$ limit. These CFT correlators will thus be dual to the worldsheet correlators that have been deformed by the worldsheet analogue of the perturbation (\ref{D}), see \cite{Fiset:2022erp}.

\section{Conclusions}\label{sec:conclusion}
In this paper we have shown that the correspondence relating tensionless string theory on 
${\rm AdS}_3 \times {\rm S}^3 /\mathbb{Z}_k \times \mathbb{T}^4$ to a subsector of the symmetric orbifold of $\mathbb{T}^4$, see \cite{Martinec:2001cf,Gaberdiel:2023dxt}, also holds on the level of the correlation functions, at least in the planar ($N\rightarrow \infty$) limit. More concretely we have studied the correlation functions of the untwisted sector vertex operators on the worldsheet sphere, and shown that they reproduce the corresponding correlation functions in the symmetric orbifold in the large $N$ limit. On the (worldsheet) sphere all intermediate states are necessarily from the untwisted sector --- this will no longer be true once one considers higher genus worldsheet correlators --- and this is mirrored by the fact that for the symmetric orbifold correlators at large $N$, only intermediate states that involve multiple of $k$-cycles appear, see the discussion in Section~\ref{sec:3.1} and Appendix~\ref{app:B}.

It would obviously be interesting to study these correlators also at subleading order in $\frac{1}{N}$. Then also twisted sector intermediate states will appear from the worldsheet perspective, and this is also what the dual CFT analysis of Appendix~\ref{app:B} predicts.  The large $N$ behaviour of the symmetric orbifold correlators in the background of the external $k$-cycle states suggests that the relation between the chemical potential of the grand canonical ensemble and the string coupling constant \cite{Eberhardt:2021jvj} has to be modified to $p^k = g_s^{-2}$, see eq.~(\ref{pgs}) in Appendix~\ref{app:Ndep}. However, given that the dictionary between the worldsheet operators and the dual CFT operators is only simple in the large $N$ limit, see the discussion in Section~\ref{sec:vertex}, a quantitative match is likely to be somewhat subtle. Furthermore,  the twisted sector states need to be inserted at $x=0$ or $x=\infty$ (since the orbifold also acts on the spacetime coordinates, see the discussion Section~\ref{sec:vertex}), and this may  introduce additional complications. It would be very interesting to explore these questions in more detail. 

\subsection*{Acknowledgements}
We thank Samir Mathur for initial collaboration and many inisightful discussions about related topics. We also thank Ji Hoon Lee, Wei Li, and Emil Martinec for useful conversations.  The work of the group at ETH is supported by a personal grant of MRG from the Swiss National Science Foundation, by the Simons Foundation grant 994306 (Simons Collaboration on Confinement and QCD Strings), as well as the NCCR SwissMAP that is also funded by the Swiss National Science Foundation.
\nopagebreak
\appendix
\section{Spherical covering maps}\label{sec sc}
In this appendix we review some aspects of spherical covering maps. We begin by observing that every spherical covering map can be written as  

\be\label{map}
x=\Gamma(z)=b\prod_{j=1}^{m}(z-r_j)^{N_j}  \ ,
\ee
where $b$ is a constant, while each $N_j$ is an integer which can be either positive or negative. Near $z= r_j$, $\Gamma(z)$ behaves as 
\begin{align}\label{lb1}
x  \approx c_j(z-r_j)^{N_j} \ ,
\end{align}
where $c_j$ is a constant.
A positive $N_j$ corresponds to an initial cycle of length $N_j$ located at $x=0$ and $z=r_j$, while a negative $N_j$ corresponds to a final cycle of length $|N_j|$ located at $x=\infty$ and $z=r_j$. 
As $z\to \infty$, the covering map is asymptotically of the form 
\be\label{lb2}
x\approx b z^{\tilde N} \qquad \hbox{with  $\ \tilde N=\sum_{j=1}^m N_j$\ .}
\ee
In the following we want to assume that $\tilde N\neq 0$; otherwise, i.e.\ if $\tilde{N}=0$, we can obtain a positive $\tilde{N}$ by applying a M\"obius transformation to the variable $z$, since we can move a preimage of $x=\infty$ to $z=\infty$. (Alternatively, we can also move a preimage of $x=0$ to $z=\infty$, which would then give rise to a negative $\tilde N$). 

In addition to the branch points at $z=r_j$ and $z= \infty$, there are further branch points $z=z_i$ 
that are characterised by the condition $\Gamma'(z_i)=0$. Near each branch point, the covering map behaves as
\begin{align}\label{xj0}
x\to x_i:&\quad x-x_i \approx a_i(z-z_i)^{w_i} \ , \qquad   i = 1,\ldots, n \ ,
\end{align}
where $w_i$ are integers greater than $1$. 

\subsection{An identity for spherical covering maps}\label{sec:identity}

For a general spherical covering map, no simple analytic formula for the locations $x_i$ of the branch points is known. However, a simple relation exists between the coefficients $c_j$ and the locations $x_i$, given by 
\be\label{i0}
\prod_{j=1}^m c_j = \frac{b \tilde N^{\tilde N}}{\prod_{j=1}^m N_j^{N_j}}\prod_{i=1}^{n}x^{w_i-1}_i \ .
\ee
In this appendix, we prove this relation; it plays an important role in the evaluation of the CFT correlator, see eq.~(\ref{Ck}). 
\smallskip

\noindent To derive (\ref{i0}) we begin by taking the derivative of $\Gamma(z)$, 
\be\label{Gder}
\Gamma'(z) = b \, \prod_{k=1}^m(z-r_k)^{N_k-1} \, \sum_{\ell=1}^{m} N_\ell \, \prod_{p\neq \ell}(z-r_p) \ .
\ee
In order to characterise the behaviour of $\Gamma(z)$ at $z=z_j$, we rearrange this to 
\begin{align}\label{i1}
\frac{\Gamma'(z)}{b\prod_{k=1}^m(z-r_k)^{N_k-1}}  \Bigg|_{z=r_j} = N_j\prod_{p\neq j}(r_j-r_p) \ ,
\end{align}
where we have used that on the right-hand-side only the term $\ell=j$ contributes. 

As above, see eq.~(\ref{xj0}), we denote the critical points of $\Gamma(z)$ (apart from those arising at $z=r_j$) by $z=z_i$, $i=1,\ldots,n$, where $z=z_i$ has multiplicity $w_i-1$. As a consequence $\Gamma'(z)$ in (\ref{Gder}) must be, up to a constant, the product of the corresponding factors. This therefore implies 
\begin{align}\label{i2}
\frac{\Gamma'(z)}{b\prod_{k=1}^m(z-r_k)^{N_k-1}}   = \tilde N \prod_{i= 1}^{n}(z-z_i)^{w_i-1} \ ,
\end{align}
where the constant prefactor $\tilde N$ can be determined from the behaviour of $\Gamma'(z)$ for $z$ large, using that $\Gamma'(z)= \tilde N b z^{\tilde N-1}+\dots$, see eq.~(\ref{lb2}). Setting $z=r_j$ in (\ref{i2}) and equating it with (\ref{i1}), we thus obtain, for each $j=1,\ldots, m$
\be\label{ii0}
N_j\prod_{p\neq j}(r_j-r_p)= \tilde N \prod_{i= 1}^{n}(r_j-z_i)^{w_i-1} \ ,
\ee
which can be written as
 \be\label{ii}
N_j\prod_{p\neq j}(r_p-r_j)= \tilde N \prod_{i= 1}^{n}(z_i-r_j)^{w_i-1} \quad \hbox{or} \quad 
N_k\prod_{j\neq k}(r_j-r_k)= \tilde N \prod_{i= 1}^{n}(z_i-r_k)^{w_i-1} \ .
\ee
Here we have used that, in going from (\ref{ii0}) to (\ref{ii}), the number of minus signs on the left-hand side, $m-1$, equals the number of minus signs on the right-hand-side, i.e.\ $\sum_{i} (w_i -1)$. This follows from the Riemann-Hurwitz formula for genus zero covering maps (of the sphere),
\be
N_{\rm active} - 1 = \frac{1}{2} \sum_{i=1}^{n} (w_i-1) + \frac{1}{2} \sum_{j=1}^{m} (|N_j|-1) + \frac{1}{2}  (|\tilde N|-1) \ ,
\ee
together with the fact that the total length of the initial and final cycles double-counts the active colours $N_{\rm active}=\frac{1}{2}(|\tilde N|+\sum_{j=1}^{m} |N_j|)$, from which we conclude
\be
\sum_{i=1}^{n} (w_i-1) = m -1  \ .
\ee
With these preparations we are now ready to prove (\ref{i0}). It follows from eq.~(\ref{lb1}) that 
\be
\prod_{j=1}^m c_j = 
b^m\prod_{j= 1}^{m} \prod_{k\neq j}(r_j-r_k)^{N_k} =b^m \prod_{k=1}^m\prod_{j\neq k} (r_j-r_k)^{N_k} \ .
\ee
For each factor we now use the second equation of eq.~(\ref{ii}) and hence find 
\begin{align}\label{i3}
\prod_{j=1}^m c_j =\ &b^m \prod_{k=1}^m\left(\frac{\tilde N}{N_k}\prod_{i= 1}^{n}(z_i-r_k)^{w_i-1}\right)^{N_k}
=\frac{b^m {\tilde N}^{\tilde N}}{\prod_{j=1}^m N_j^{N_j}}\prod_{i= 1}^{n}\prod_{k=1}^m(z_i-r_k)^{N_k(w_i-1)} \n
=\ & \frac{b {\tilde N}^{\tilde N}}{\prod_{j=1}^m N_j^{N_j}}\prod_{i= 1}^{n} \big(\Gamma(z_i)\big)^{w_i-1} =\frac{b {\tilde N}^{\tilde N}}{\prod_{j=1}^m N_j^{N_j}}\prod_{i= 1}^{n} x_i^{w_i-1} \ ,
\end{align}
thus proving (\ref{i0}).

\subsection{Correlator from spherical covering map}\label{app c}

The symmetric orbifold correlator can be evaluated by a sum over all covering maps, where each covering map contributes the conformal factor associated to the covering map, times the correlator of the seed theory on the covering surface. In this paper we are only considering sphere coverings --- these are the leading contributions at large $N$ --- and for them a compact formula for the corresponding contribution was derived in \cite{Lunin:2000yv,Dei:2019iym}. In the following we will need a slight generalisation of their results, see below.  We claim that in the general case (\ref{map}), the conformal factor from the covering map equals 
\begin{align}\label{fr0a}
C_{\Gamma}
= |\tilde N|^{\frac{1}{2}(\tilde N+1)}|b|^{\frac{1}{2}(1-\frac{1}{\tilde N})}\Big(\prod_{j=1}^m |N_j|^{-\frac{1}{2}(N_j+1)} |c_j|^{-\frac{1}{2}(1-\frac{1}{N_j})}\Big)
\Big(\prod_{i=1}^n w_i^{-\frac{1}{2}(w_i+1)} |a_i|^{-\frac{1}{2}(1-\frac{1}{w_i})}\Big) \ ,
\end{align}
and the correlation function simply becomes 
\be\label{vca}
\big\langle  \prod_{i=1}^{n} V_i( x_{i,R}) \big\rangle = C_{\Gamma}\, \big\langle  \prod_{i=1}^{n} \tilde V_i( z_{i})  \big\rangle \ ,
\ee
where the operators with a tilde represent the corresponding operators on the covering surface.
Notice that the correlation function depends only on the twist orders $N_j$, $\tilde N$, and $w_i$, as well as the prefactors $c_j$, $b$, and $a_i$ introduced in (\ref{lb1}), (\ref{lb2}), and (\ref{xj0}). The dependence on the twist locations $x_i$ is encoded implicitly via the above parameters.  
\smallskip

In \cite{Lunin:2000yv,Dei:2019iym}, the contribution from the spherical covering map (\ref{map}) was only provided for $N_j\geq -1$. In this appendix, we present the explicit result for arbitrary $N_j$, including the cases where $N_j\leq -2$, which correspond to higher-order poles of the covering map (or a branch point at $x=\infty$ and finite $z$). These contributions can be obtained by applying the conformal transformation $\tilde x= -\frac{1}{x}$, which maps $N_j\leq -2$ to $N_j\geq 2$. Below, we verify that the correlator (\ref{fr0}) behaves correctly under this transformation.

Under the conformal transformation, the covering map (\ref{map}) becomes
\be\label{conf}
\tilde x = -\frac{1}{x} = -b^{-1}\prod_{j=1}^{m}(z-r_j)^{-N_j} \ . 
\ee
The branch points in (\ref{xj0}) become
\begin{align}\label{xj1}
\tilde x\to -\frac{1}{x_i}:&\quad \tilde x-\big(-\frac{1}{x_i}\big) \approx \frac{a_i}{x_i^2}(z-z_i)^{w_i} \ .
\end{align}
Using the result (\ref{fr0}), the correlation function after the transformation is given by making the replacements $N_j\to -N_j,\, \tilde N\to -\tilde N, \, c_j\to - c_j^{-1}, \,  b\to -b^{-1}, \,  a_i \to \frac{a_i}{x_i^2}$, which gives
\begin{align}\label{fr1}
\tilde C_{\Gamma}
=\ &|\tilde N|^{\frac{1}{2}(-\tilde N+1)}|b|^{-\frac{1}{2}(1+\frac{1}{\tilde N})}\Big(\prod_{j=1}^m |N_j|^{-\frac{1}{2}(-N_j+1)} |c_j|^{\frac{1}{2}(1+\frac{1}{N_j})}\Big) \n
&\quad \times \Big(\prod_{i=1}^n w_i^{-\frac{1}{2}(w_j+1)} \Big|\frac{a_i}{x_i^2}\Big|^{-\frac{1}{2}(1-\frac{1}{w_i})}\Big) \ .
\end{align}
Thus, we find
\be
\tilde C_{\Gamma}/C_{\Gamma}= \left|\frac{b^{-1} \tilde N^{-\tilde N}}{\prod_{j=1}^m N_j^{-N_j}}\prod_{j=1}^m c_j
\prod_{i=1}^n x_i^{1-\frac{1}{w_i}}\right| \ . 
\ee
Together with the identity (\ref{i0}), this can then be rewritten as 
\be\label{tc/c}
\tilde C_{\Gamma}/ C_{\Gamma}= 
\prod_{i=1}^n\left|x_i\right|^{w_i-\frac{1}{w_i}} \ .
\ee
Let us now verify that this agrees with conformal symmetry.
Under the transformation (\ref{conf}), an operator of dimension $h,\bar h $ transforms as 
\be\label{trans}
O(-\tfrac{1}{x},-\tfrac{1}{\bar x}) = x^{2h} \bar x^{2\bar h} O(x,\bar x) \ .
\ee
The twist operator corresponding to the branch point (\ref{xj0}) has dimension $h=\bar h = \frac{1}{4}(w_i-\frac{1}{w_i})$, which gives
\be
O(-\tfrac{1}{x_i},-\tfrac{1}{\bar x_i})= \left|x_i\right|^{w_i-\frac{1}{w_i}}O(x_i,\bar x_i) \ .
\ee
This accounts for the factor appearing in (\ref{tc/c}). Note that operators located at $x=0$ and $x=\infty$ do not contribute additional factors, as operators at infinity are defined via a scaling $O(\infty)= \lim_{x\to \infty} x^{-2h}\bar x^{-2\bar h}  O(x,\bar x)$ for which the prefactor cancels the factor in (\ref{trans}).

\section{Permutation structure of intermediate states}\label{app:B}

In this appendix we give an explicit symmetric group argument for the claim that the covering map in question is the $k$'th power of another covering map, see eq.~(\ref{key}) in Section~\ref{sec:3.1}.

Let us consider the action of a twist operator $\sigma$ of length $w$ on an initial state that consists of a product of cycles,
\be\label{te}
(X)_\sigma\Big[\prod_{i=1}^{L^{in}}(Y_i)\Big] = \Big[\prod_{i=1}^{L^{out}}(\tilde Y_i)\Big] \ , \qquad X\subset \bigcup_{i=1}^{L^{in}} Y_i = \bigcup_{i=1}^{L^{out}} \tilde Y_i \ .
\ee
Here, upper case letters $X$ and $Y$ represent sets of distinct numbers within the corresponding cycle, while lower case letters $a$ and $b$ will be used later to denote the individual numbers within a cycle. So, for example, $X$ contains the $w$ numbers that make up the $w$-cycle permutation $\sigma$. Furthermore, we need to include all the cycles in the initial state that contain a number appearing in $X$ --- they are all affected by the multiplication by $\sigma$ --- and hence $X$ is contained in the union of the $Y_i$. Similarly, we need to keep track of all of these colours in writing the out-state, and therefore $X$ is also contained in the union of the $\tilde{Y}_i$. 

We denote by $w^{(in)}_i = |Y_i|$  the length of the corresponding cycle, and similarly for $w^{(out)}_i = |\tilde{Y}_i|$. (With this convention we then have $|X|=w$.)

\subsection{Bounding the total number of cycles}

The first step consists of showing that the total number of initial and final cycles, $L^{in}$ and $L^{out}$ in (\ref{te}), satisfies the bound
\be\label{ml}
L^{in}+L^{out}\leq w+1 \ .
\ee
This bound can be directly derived from the Riemann-Hurwitz formula
\be\label{RH}
N_{\rm active} = 1- g + \sum_{i} \frac{w_i - 1}{2} \ ,
\ee
where $N_{\rm active}$ is the total number of `active colours', i.e.\ the numbers involved in the product, $g\geq 0$ is the genus of the corresponding covering space, and $w_i$ denotes the twist order of each twist operator, i.e.\ the different $w_i$ are $w$, $w^{(in)}_i$, and $w^{(out)}_i$.

To derive (\ref{ml}), we first observe that because of the second equation in (\ref{te}), 
the number of active numbers is given by
 \begin{align}\label{Nw}
 N_{\rm active}=\sum_i w^{(in)}_i=\sum_j w^{(out)}_j \ .
 \end{align}
Thus we find 
 \begin{align}\label{sw}
 \sum_{i} \frac{w_i-1}{2} = N_{\rm active} + \frac{1}{2}\big(  w -L^{in}-L^{out} -1\big)
 \ .
 \end{align}
Substituting this into the Riemann-Hurwitz formula (\ref{RH}) and using that $g\geq 0$, we arrive at eq.~(\ref{ml}).\footnote{At this stage it is maybe not obvious that we can use the Riemann-Hurwitz formula since this refers to the entire correlator, whereas here we are only looking at the application of a single twist field on a reference state. However, it will become clear from the arguments in Section~\ref{app:B2}, see the paragraph above eq.~(\ref{B.15}), that this is the correct condition also at this intermediate stage.}

Next we want to find an explicit description of the relevant permutations that saturate this bound. To do so, we consider recursively moves that allow us to simplify the expression in question. First, let us assume that $\sigma$ contains two adjacent colours that appear in one cycle in the initial state, i.e.\ a configuration of the form 
\be\label{B.6}
(X\,a\,b)_\sigma \Big[(b\,Y a\,Y')\prod_i(Y_i)\Big] 
=(b\,Y)\, (X a)_\sigma \Big[(a\,Y')\prod_i(Y_i)\Big] \, , \quad X\cap Y=\emptyset \ , 
\ee 
where in addition we have assumed that the other colours of $\sigma$, i.e.\ $X$, do not appear in $Y$. To see the equality in (\ref{B.6}) we note that the last number of $Y$ gets permuted as $\text{last}(Y) \to a \to b$, and hence the segment $(b\, Y)$ forms a separate final cycle by itself, which we can move to the very left since $X\cap Y=\emptyset$. Similarly, the last number of $Y'$ gets permuted as $\text{last}(Y') \to b \to \text{First}(X) $ on the left-hand side, and as $\text{last}(Y') \to a \to \text{First}(X) $ on the right-hand side. 

We want to think of this operation as a `move' by means of which we can simplify the original product of permutations. In particular, we can remove the segment $(b\, Y)$ from both sides of the above identity, and hence reduce it to the configuration
\be
(X\,a\,b)_\sigma \Big[(b\,Y a\,Y')\prod_i(Y_i)\Big] \
\Rightarrow \ (X a)_\sigma  \Big[(a\,Y')\prod_i(Y_i)\Big] \, , \quad X\cap Y=\emptyset \ .
\ee
This operation reduces both the order of the twist, and the number of final cycles by one
\be
w\to w -1 \ , \qquad L^{out} \to L^{out}-1 \ ,
\ee
and hence affects both sides of the inequality (\ref{ml}) in the same way.

The second move involves a configuration where $\sigma$ only shares a single number with one of the initial cycles, i.e.\ a configuration of the form 
\be\label{so1}
(Xa\,b)_\sigma\Big[(a\, Y)\prod_i(Y_i)\Big] = (b\,a\,Y)\, (X\, b)_\sigma \Big[\prod_i(Y_i)\Big] \ ,
\ee
where $b\not\in Y$ and $X\cap Y=\emptyset$. 
This allows us to remove the segment $(a\, Y)$ from both sides of the above identity, and hence 
reduce it to the configuration
\be\label{so}
(Xa\,b)_\sigma\Big[(a\, Y)\prod_i(Y_i)\Big] \  \Rightarrow \  (X\, b)_\sigma \Big[\prod_i(Y_i)\Big] \ .
\ee
Note that the cycle $(b\,a\,Y)$ in (\ref{so1}) only shares the number $b$ with the remaining cycles on which it acts; as a consequence the number of final cycles is not modified. However, the above move reduces $w$ and $L^{in}$ by one, i.e.\ 
\be
w\to w -1 \ , \qquad L^{in} \to L^{in}-1 \ ,
\ee
and hence affects both sides of the inequality (\ref{ml}) in the same way. 

By repeatedly applying these two operations, we can systematically simplify the action of a twist operator while preserving the condition (\ref{ml}). In this way, we can reduce any configuration to either the `trivial' case, for which $w=1$, i.e.\ 
\be\label{tta}
(a)_\sigma\big[(a\, Y)\big] = \big[(a\, Y)\big]  \ .
\ee
Alternatively, the process may lead to an action where each initial cycle shares two or more numbers with the twist operator, and all final cycles contain at least two segments $Y_i$,\footnote{Otherwise, we can use the  first move to remove that final cycle.} e.g.\ examples of the kind 
\begin{align}\label{nta}
(abcd)_\sigma \big[(a\,Y_1\,  c\,Y_2)(b\,Y_3 \, d\,Y_4)\big] &= \big[ (a \, Y_1 \, d \, Y_4 \, c \, Y_2 \, b \, Y_3)\big] \ , \n
(abc)_\sigma \big[ (a \, Y_1\, b \, Y_2 \, c \, Y_3) \big] &=\big[ (a \, Y_1\, c \, Y_3 \, b \, Y_2)\big] \ .
\end{align}
We now want to show that one can saturate the condition (\ref{ml}) if and only if we can reduce the configuration to the trivial case by the above two moves. Obviously, since the trivial configuration (\ref{tta}) saturates the bound ($w=1$ and $L^{(in)}=L^{(in)}=1$), this proves the `if' part of the above statement. So let us assume that this is not the case, i.e.\ that the final configuration has the property that each initial cycle shares two or more numbers with the twist operator, and that all final cycles consist of at least two segments. The first condition implies that $L^{(in)}\leq w/2$, and the second condition implies $L^{(out)}\leq w/2$ --- the number of segments equals the order of the twist (as the segments are separated by the members from the twist). But then it is clear that (\ref{ml}) is not saturated. 

 Thus it follows that the general structure of a twist action that saturates (\ref{ml}) can be reconstructed from the trivial twist action by reversing the two moves; it therefore has the general form 
\begin{align}\label{gta}
(12\dots w)_\sigma \Big[\prod_{i=1}^m \big(\prod_{j=1}^{n_i} a_{i,j}Y_{i,j}\big) \Big] 
=  \big(\prod_{i=1}^m a_{i,n_i}Y_{i,n_i}\big)\prod_{i=1}^m \prod_{j=1}^{n_i-1}\big( a_{i,j}Y_{i,j}\big)  \ .
\end{align}
Here $a_{i,j}\in \{1,\dots , w\}$, and the numbers appear in reverse order, i.e.\ $a_{i,j}>a_{k,l}$ if $i<k$ or $i=k$ and $j<l$. In the resulting cycles, all the segments $a_{i,j}Y_{i,j}$ form separate cycles, except for the last segment $a_{i,n_i}Y_{i,n_i}$ in each initial cycle. All of these cycles then join into a final cycle $\big(\prod_i a_{i,n_i}Y_{i,n_i}\big)$. This twist action saturates (\ref{ml}), where the number of initial cycles is $m$ and the number of final cycles is $w-m+1$.

\subsection{Spherical covering maps}\label{app:B2}

\begin{figure}[h]
\centering
        \includegraphics[width=2cm]{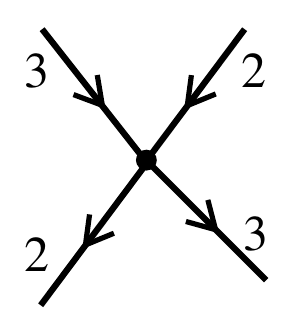}
\caption{The diagram representing $(234)_\sigma\big[(123)(45)\big] = (13)(245)$.}
\label{fig3}
\end{figure}

It is convenient to describe these special covering maps diagrammatically: for each twist operator we introduce a vertex, and  initial and final cycles are represented as incoming and outgoing lines with arrows. The `momentum' of each line corresponds to the cycle length. It is evident that the length is conserved at each twist operator. For example, a twist-3 operator with the action $(234)_\sigma [ (123)(45)]=(13)(245)$ is shown in Fig.~\ref{fig3}, with two incoming lines of lengths $3$ and $2$, and two outgoing lines of lengths $2$ and $3$. While this diagram omits some details about how the cycles are connected, it helps clarify the twist structure in the large $N$ expansion.

\begin{figure}[h]
\centering
        \includegraphics[width=4cm]{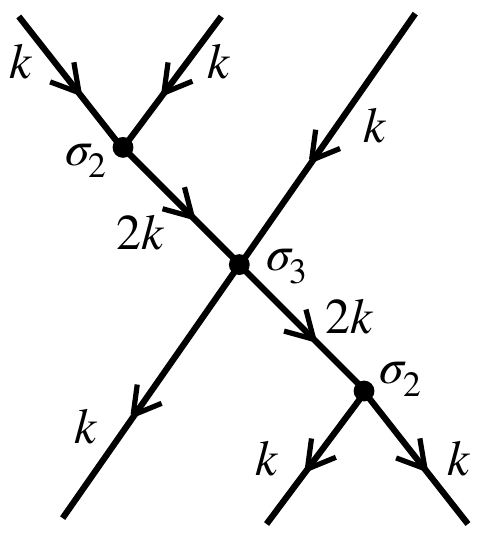}
\caption{A tree-like diagram for $\sigma_2\sigma_3\sigma_2$ with three initial and final cycles of length $k$.}
\label{fig9}
\end{figure}

In a correlation function involving multiple twist operators, the final cycle of one twist operator can serve as the initial cycle for another, and vice versa. This allows us to connect the lines from the twist operators to form a diagram. For example, in Fig.~\ref{fig9}, we show the correlation function involving three twist operators $\sigma_2\sigma_3\sigma_2$ and three initial and final cycles, each of length $k$. In this diagram, all the twist operators saturate the condition (\ref{ml}).

It is straightforward to observe that to maximise the total number of initial and final lines, the diagram must be tree-like, with each twist operator saturating the condition (\ref{ml}). In this configuration, the total number of initial and final lines is given by
\begin{align}\label{B.15}
\text{max}(L^{in}+L^{out}) = \sum_{i}(w_i-1) + 2 \ .
\end{align}
This result can be derived by successively adding twist operators. For one twist operator, the result is $w_1+1$, and each subsequent twist operator increases the total number by $w_i-1$. This result can also be obtained by using the Riemann-Hurwitz formula, following a similar procedure as described around eq.~(\ref{Nw}).

Now, consider the setup described in this paper, where all initial and final cycles have length $k$. For a spherical covering map, the number of initial and final cycles is maximised. As discussed earlier, this implies that the diagram must be tree-like. Moreover, since the cycle length is conserved at each twist operator, the lengths of the intermediate cycles are fully determined and are all multiples of $k$.\footnote{Conversely,  
 for a diagram with a loop, the lengths of the intermediate cycles are not fully determined and are not required to be multiples of $k$.} Note that the fact that all intermediate states only involve cycle lengths that are a multiple of $k$ is crucial for writing the covering map as the $k$'th power of another covering map (where we collapse all $wk$-cycles to $w$-cycles). 

We also note that for a spherical covering map, the numbers in any twist operator are separated by multiples of $k$ in the intermediate states, where the separation between the $j$-th and $i$-th numbers in a cycle is defined as $|j-i|$. As discussed earlier, for a spherical covering map, all the twist operators saturate condition (\ref{ml}), and therefore, their actions are given by (\ref{gta}). We now need to show that all segments $a_{i,j}Y_{i,j}$ have lengths that are multiples of $k$. We note that all of these segments, except for the case $j=n_i$, appear as individual cycles on the left hand side of (\ref{gta}), and thus, their lengths must be multiples of $k$, as they correspond to intermediate or final cycles. For the case where $j=n_i$, since all the initial cycles on the right hand side of (\ref{gta}) have lengths that are multiples of $k$, by excluding the segments where $j\neq n_i$, which already have lengths that are multiples of $k$, the remaining segment $a_{i,n_i}Y_{i,n_i}$ must also have a length that is a multiple of $k$. This demonstrates the claim.

\section{Large $N$ expansion}\label{app:Ndep}

In this appendix we explore the large $N$ expansion of the $n$-point correlators in the presence of the $k$-cycle external states, see eq.~(\ref{aim}). We begin by recalling the situation for the case where all initial and final cycles are identical 1-cycles; this was studied in \cite{Lunin:2000yv}, see also \cite{Pakman:2009zz}. The result takes the form
\be
N^{s-\sum_{i}\frac{w_i}{2}} = N^{-(g+\frac{n-2}{2})} \ ,
\ee
where $s$ denotes the number of active colours and $g$ is the genus of the covering surface. It shows that the large $N$ expansion naturally organises itself into a genus expansion. 

In this appendix, we generalise the analysis to the case of interest, where the initial and final cycles are identical $k$-cycles. As we shall see below, see eq.~(\ref{lc}), in the large $N$ limit, the leading contribution arises from spherical covering maps that maximise the number of active $k$-cycles, while higher genus contributions are suppressed by powers of $N^{-k}$. (This suppression can also be understood by noting that at each higher subleading order, the number of active colours involved decreases by $k$.) In particular, this therefore suggests that the dictionary between the string coupling constant and the chemical potential of the grand canonical partition function of \cite[eq.~(5.20)]{Eberhardt:2021jvj} needs to be modified for the $\mathbb{Z}_k$ orbifold to 
\be\label{pgs}
p^k = g_{s}^{-2} \ .
\ee
(This is the relation provided that the dual CFT is evaluated on the sphere, i.e.\ $G=0$ in \cite[eq.~(5.20)]{Eberhardt:2021jvj}.)

Let us begin with a background state consisting of $N/k$ identical $k$-cycle states, with the total number of colours being $N$. We denote this state by $\Psi$. In the symmetric orbifold theory with permutation group $S_{N}$, any operator associated to a single cycle permutation of length $w$ needs to be averaged over the entire conjugacy class of $w$-cycles
\be
V_{w}= \frac{\lambda_w}{N!}\sum_{h\in S_{N}} \sigma_{h(1\cdots w)h^{-1}} \ ,
\ee
where $\sigma_g$ denotes the twist-operator corresponding to the permutation $g$. Next we consider the correlator of $n$ such operators acting on the background state $\Psi$
\begin{align}\label{n1}
\langle  \Psi | V_{w_1}\cdots V_{w_n}|\Psi \rangle
&= \left(\frac{1}{N!}\right)^n \lambda_{w_1}\dots  \lambda_{w_n} 
\frac{N!}{(sk)!(N-sk)!}(N-w_1)!\dots (N-w_n)! \n
 &\qquad\times  \sum_{1\leq j_{k,m_k}\leq sk}\langle\Psi| \sigma_{(j_{1,1}\dots j_{1,w_1})} \dots \sigma_{(j_{n,1}\dots j_{n,w_n})}|\Psi \rangle  \n
 &\propto N^{ s k-\sum_i w_i } \lambda_{w_1}\dots  \lambda_{w_n} \ .
\end{align}
Here, the active colours are labeled by $1,\ldots, sk$, corresponding to $s$ active $k$-cycles. The sum is over all possible choices within these active colours and therefore does not introduce any additional $N$-dependence. The combinatorial factor $\frac{N!}{(sk)!(N-sk)!}$ counts the number of ways of choosing $sk$ active colours from the total of $N$ colours, while the factor $(N-w_i)!$ arises from permutations that leave the colours $(j_{i,1}\dots j_{i,w_i})$ of $\sigma_{w_i}$ untouched. Note that the number of active $k$-cycles $s$ is bounded above by half of (\ref{B.15})
\be
s\leq 1+\sum_{i}\frac{w_i-1}{2} \ .
\ee
To determine the normalisation constant $\lambda_w$, we consider the 2-point correlator
\be
\langle \Psi | V_{w}\, V_{w} |\Psi \rangle  \propto N^{ ( k-2)w} \lambda^2_w \ .
\ee
Here, the number of active $k$-cycles is taken to be its maximal value $w$. This implies that the normalisation constant must scale as
\be
 \lambda_w \propto  N^{ (1-\frac{k}{2})w} \ .
\ee
Substituting this into the correlator (\ref{n1}), we find
\be\label{lc}
\langle\Psi| V_{w_1}\dots V_{w_n}|\Psi \rangle_k \propto N^{  k(s-\sum_i \frac{w_i}{2})} = N^{  -k(g+ \frac{n-2}{2})} \ ,
\ee
where $s$ is the number of active $k$-cycles, and $g$ is the genus of the covering surface.
The final expression follows from the Riemann-Hurwitz formula
\be\label{RH1}
sk = 1- g + \sum_{i=1}^n \frac{w_i - 1}{2} + 2s\frac{k - 1}{2} \ ,
\ee
where the last term accounts for the contributions from the $2s$ initial and final $k$-cycles.

\end{document}